\documentclass[twocolumn,aps,epsfig,nofootinbib,floatfix]{revtex4}

\usepackage{graphicx}
\usepackage{epstopdf}
\usepackage{latexsym}
\usepackage{amssymb}
\usepackage{amsmath}
\usepackage{mathrsfs}
\usepackage{wasysym}
\usepackage{textcomp}
\usepackage{braket}
\usepackage{color}
\usepackage{epstopdf}
\usepackage[center]{subfigure}

\begin{document}

 \newcommand{\bq}{\begin{equation}}
 \newcommand{\eq}{\end{equation}}
 \newcommand{\bqn}{\begin{eqnarray}}
 \newcommand{\eqn}{\end{eqnarray}}
 \newcommand{\nb}{\nonumber}
 \newcommand{\lb}{\label}
\newcommand{\PRL}{Phys. Rev. Lett.}
\newcommand{\PL}{Phys. Lett.}
\newcommand{\PR}{Phys. Rev.}
\newcommand{\CQG}{Class. Quantum Grav.}
 \newcommand{\sst}{\scriptscriptstyle}
 \newcommand{\lrp}[1]{\left(#1\right)}
 \newcommand{\lrb}[1]{\left[#1\right]}
 \newcommand{\lrc}[1]{\left\{#1\right\}}
 \newcommand{\hongse}[1]{\textcolor{red}{#1}}
\title{Effects of parity violation on  non-gaussianity of primordial  gravitational waves in Ho\v{r}ava-Lifshitz  gravity}

\author{Tao Zhu $^{a,  b}$}
\email{Tao_Zhu@baylor.edu}

\author{Wen Zhao $^{c}$}
\email{wzhao7@mail.ustc.edu.cn}

\author{Yongqing Huang $^{a}$}
\email{Yongqing_Huang@baylor.edu}

\author{Anzhong Wang $^{a}$}
\email{Anzhong_Wang@baylor.edu}

\author{ Qiang Wu $^{b}$}
\email{wuq@zjut.edu.cn}

\affiliation{$^{a}$ GCAP-CASPER, Physics Department, Baylor University, Waco, TX 76798-7316, USA\\
$^{b}$  Institute for Advanced Physics $\&$ Mathematics, Zhejiang University of Technology, Hangzhou, 310032, China\\
$^{c}$ Department of Astronomy, University of Science and Technology of China, Hefei, 230026, China }

\date{\today}

\begin{abstract}

In this paper, we study the effects of parity violation on non-gaussianities of primordial gravitational waves in the framework
of Ho\v{r}ava-Lifshitz theory of gravity, in which high-order spatial derivative operators, including the ones violating parity, 
generically appear. By calculating the three point function, we find that the leading-order contributions to the non-gaussianities 
come from the usual second-order derivative terms, which produce the same bispectrum as that found in general relativity. 
The contributions from high-order spatial  n-th derivative terms are always suppressed by a factor $(H/M_*)^{n-2} \; (n \ge 3)$, 
where $H$ denotes the inflationary energy and $M_*$ the suppression mass scale of  the high-order spatial derivative 
operators of the theory. Therefore, the next leading-order contributions come from the 3-dimensional gravitational Chern-Simons 
term. With some reasonable arguments, it is shown that this 3-dimensional operator is  the only one that violates the parity and 
in the meantime  has non-vanishing contributions to non-gaussianities.

\end{abstract}

\pacs{ 98.80.Cq, 98.80.-k, 04.50.-h}

\maketitle
\section{Introduction}
\renewcommand{\theequation}{1.\arabic{equation}} \setcounter{equation}{0}

Primordial gravitational waves (PGWs), which are expected to be generated during inflation, have attracted a great deal of attention recently, as their detections would be the direct evidence of inflation, and more important  the existence of gravitational waves in the universe. From the properties of PGWs, such as their power spectra and non-gaussianities, we can extract useful information about the theory of inflation and gravity. In particular,  the PGWs produce not only the temperature anisotropy, but also a distinguishable  signature in the polarization of the cosmic microwave background (CMB)  \cite{KDM}. Decomposing the polarization into two modes: one is curl-free, the E-mode, and the other is divergence-free, the B-mode, one finds that the B-mode pattern cannot be produced by density fluctuations. Thus, its detection would provide a unique signature for the existence of PGWs \cite{seljak}. 

In addition, PGWs normally produce the TT, EE, BB and TE spectra of CMB, but the spectra of TB and EB vanish when  the parity of the PGWs  is conserved \cite{KDM}.  However, if the theory is chiral, the power spectra of right-hand and left-hand PGWs can have different amplitudes, and then induce non-vanishing TB and EB correlation in  large scales \cite{lue}. This provides the opportunity to directly detect the chiral asymmetry of the theory by observations  \cite{lue,saito,kamionkowski}.  Recently, in \cite{soda,WWZZ} the above mentioned  problem was addressed   in the  framework of Ho\v{r}ava-Lifshitz (HL) theory of gravity \cite{Horava}, in which the Lorentz symmetry is broken in the ultraviolet (UV), and parity-violating operators generically appear. In particular, it was  shown that,  because of the parity violation and non-adiabatic evolution of the modes, a large polarization of PGWs can be produced, and  could be well within the range of detection of the forthcoming CMB observations \cite{WWZZ}. 

The effects of  the parity violation on  non-gaussianities of PGWs were also studied \cite{parity,parity-soda} in the theory with the general covariance, and  shown that, because of the symmetry of the pure de Sitter background, the parity violation from Weyl cubic terms have no contributions to the non-gussianities, although this is no longer true when the  coupling of Weyl cubic terms is time-dependent  \cite{parity2}. It should be noted that in all these studies the symmetry of   the general diffeomorphisms of the underlaid theories  plays an crucial role. On the other hand, in the HL theory the symmetry  is reduced to the foliation-preserving diffeomorphisms \cite{Horava}, and the parity-violating  operators allowed by such a symmetry are quite different from those with the general diffeomorphisms. Thus, it is expected that  in the HL theory some distinguishable features  of non-gaussianities of PGWs due to these parity-violating operators should exist, which may provide a smoking gun for the tests of the HL theory in the forthcoming CMB observations.  With these motivations, in this paper we study the non-gaussianities of PGWs in the HL theory, and focus ourselves  mainly on the effects of the high-order spatial operators on the non-gaussianities of PGWs, especially on the ones that violate  the parity.

The rest of the paper is organized as follows: In Sec. II we first give a very brief review on the HL theory, and then restrict ourselves to the model recently proposed in \cite{ZWWS,ZSWW}, where an extra U(1) symmetry is enforced in the nonprojectable case, in order to eliminate the spin-0 gravitons usually appearing in the HL theory. In this section, we also present   the linearized  equation of motion of the tensor perturbations, originally derived in \cite{WWZZ}.
In Sec. III, from the cubic action of tensor perturbations, we calculate the three-point correlation function and obtain the bispectrum of the PGWs, while in Sec IV, we plot the shapes of bispectrum produced by both the second-order derivative operators and the three-dimensional parity-violating Chern-Simons one. In Sec V, we summarize our main results. There are also two appendices, A and B,  in which the cubic action is given explicitly. 

Before processing further, we would like to note that, although in this paper we restrict ourselves only to the model of the HL theory proposed recently in \cite{ZWWS,ZSWW}, our results can be easily generalized to other models, as the tensor perturbations are quite similar in all of these models \cite{WangTensor,WWZZ}.  In addition, non-gaussianities of PGWs in the framework of the general covariant theory with the projectability condition was also studied in \cite{HWYZ}, and several remarkable features were found. In particular, it was found that the  terms $R_{ij}R^{ij}$ and 
$\left(\nabla^{i}R^{jk}\right)\left(\nabla_{i}R_{jk}\right)$ exhibit a peak at the squeezed limit, while the one $R^i_j R^j_k R^k_i$ favors 
the equilateral shape when spins of the three tensor fields are the same,  but peaks in between the equilateral and squeezed limits 
when spins are mixed, where $R_{ij}$ denotes the 3-dimensional Ricci tensor made of the 3-dimensional metric $g_{ij}$ of the leaves $t = $ Constant, and $\nabla_i$ denotes the covariant derivative with respect to $g_{ij}$.   
The consistency with  the recently-released Planck observations  \cite{planck_2013}  was  also discussed. However,  in \cite{HWYZ} the parity-violating operators were excluded. 
Therefore, in this paper we shall focus mainly on the effects of these operators on non-gaussianities, as mentioned above. Moreover, non-gaussianities of scalar perturbations were also studied in the framework of the HL theory, one in the curvaton scenario  \cite{IKM} and the other in inflationary model \cite{HW12}, and some remarkable features were obtained.


\section{Nonprojectable General Covariant HL Gravity and Linear  Tensor Perturbations}
\renewcommand{\theequation}{2.\arabic{equation}} \setcounter{equation}{0}

By construction, the HL theory is power-counting renormalizable \cite{Horava}. This is achieved by breaking the symmetry of the general covariance in the UV, 
and include   only high-order spatial derivative operators, so that it remains  also unitary, a problem that has been facing for a long time  in the quantization of
gravity \cite{Stelle}. In the low energy, low dimensional operators take over, and it is expected that the Lorentz symmetry is ``accidentally" restored \cite{reviews}.
Since  Ho\v{r}ava first proposed it in 2009, the theory has attracted a lot of attention,  partially because of various remarkable features when applied to cosmology \cite{cosmology}, 
and partially because of some challenging questions, such as ghosts,  instability and strong coupling. To overcome these questions, various 
models have been proposed  \cite{reviews}, including the ones with an additional local $U(1)$ symmetry\cite{HMT,ZWWS,ZSWW}, in which  the problems 
mentioned above can be avoided by properly choosing the coupling constants appearing in the theory. 
Since in all of those models, the tensor  perturbations are almost the same \cite{WangTensor,WWZZ,HWYZ},  without loss of the generality, 
in this paper we shall work with the model  proposed in \cite{ZWWS,ZSWW}. 

\subsection{Action of the Nonprojectable General Covariant HL Gravity}

The  fundamental variables in the   nonprojectable general covariant HL gravity  proposed in \cite{ZWWS,ZSWW} are
$$
(N, N^i, g_{ij}, A, \varphi), 
$$
where $N$ and $N^i$ denote, respectively, the lapse function and shift vector in the   Arnowitt-Deser-Misner  (ADM)  decompositions
 \cite{ADM}, and  ${A}$ and ${\varphi}$ are, respectively, the $U(1)$ gauge field and Newtonian prepotential \cite{HMT}. 
Then, the corresponding total action can be cast in the form,
\bqn
\label{action}
S&=&\zeta^2 \int dtd^3x \sqrt{g} N \Big({\cal{L}}_{K}-{\cal{L}}_{V} + {\cal{L}}_{A}
+ {\cal{L}}_{\varphi}\nb\\
&& ~~~~~~~~~~~~~~~~~~~~~~~~~  + {\zeta^{-2}} {\cal{L}}_M\Big),
\eqn
where $\zeta^2 = 1/(16\pi G)$ with $G$ being the Newtonian constant, ${\cal{L}}_M$ describes matter fields, and
\bqn
\lb{2.2}
{\cal{L}}_{K} &=& K_{ij}K^{ij} - \lambda K^2,\nb\\
{\cal{L}}_{V} &=&  {\cal{L}}_{V}^{R} + {\cal{L}}_{V}^{a},\nb\\
{\cal{L}}_{A} &=& \frac{A}{N}\left(2\Lambda_g - R\right),\nb\\
{\cal{L}}_{\varphi} &=&  \varphi{\cal{G}}^{ij}\big(2K_{ij}+\nabla_i\nabla_j\varphi+a_i\nabla_j\varphi\big)\nb\\
& & +(1-\lambda)\Big[\big(\Delta\varphi+a_i\nabla^i\varphi\big)^2\nb\\
&&  ~~~~~~~~~~~~ ~ +2\big(\Delta\varphi+a_i\nabla^i\varphi\big)K\Big]\nb\\
& & +\frac{1}{3}\hat{\cal G}^{ijlk}\Big[4\left(\nabla_{i}\nabla_{j}\varphi\right) a_{(k}\nabla_{l)}\varphi \nb\\
&& ~~~~~~~~~~~ ~ + 5 \left(a_{(i}\nabla_{j)}\varphi\right) a_{(k}\nabla_{l)}\varphi\nb\\
&&
+ 2 \left(\nabla_{(i}\varphi\right)a_{j)(k}\nabla_{l)}\varphi
+ 6K_{ij} a_{(l}\nabla_{k)}\varphi\Big],
\eqn
with $\Delta \equiv \nabla^2$, and
\bqn
\lb{2.2b}
K_{ij} &=& \frac{1}{2N}\left(- \dot{g}_{ij} + \nabla_{i}N_{j} +  \nabla_{j}N_{i}\right),\nb\\
a_{i} &=& \frac{N_{,i}}{N},\;\;\; a_{ij} = \nabla_{j} a_{i},\nb\\
\hat{\cal G}^{ijlk} &=& g^{il}g^{jk} - g^{ij}g^{kl}, \nb\\
{\cal{G}}_{ij} &=& R_{ij} - \frac{1}{2}g_{ij} R + \Lambda_g g_{ij},\nb\\
{\cal{L}}_V^R&=&\gamma_0 \zeta^2 + \gamma_1 R+\frac{\gamma_2 R^2+\gamma_3 R_{ij}R^{ij}}{\zeta^2}+\frac{\gamma_5}{\zeta^4} C_{ij}C^{ij},\nb\\
{\cal{L}}_{V}^{a} &=&    -  \beta_0  a_{i}a^{i}
  + \frac{1}{\zeta^{2}}\Bigg[\beta_{1} \left(a_{i}a^{i}\right)^{2} + \beta_{2} \left(a^{i}_{\;\;i}\right)^{2} \nb\\
& & + \beta_{3} \left(a_{i}a^{i}\right)a^{j}_{\;\;j} + \beta_{4} a^{ij}a_{ij} + \beta_{5}
\left(a_{i}a^{i}\right)R \nb\\
& &   + \beta_{6} a_{i}a_{j}R^{ij} + \beta_{7} Ra^{i}_{\;\;i}\Bigg]  +  \frac{1}{\zeta^{4}}\  \beta_{8} \left(\Delta{a^{i}}\right)^{2}.
\eqn
Here $R$ denotes the Ricci scalar, and $C_{ij}$ the Cotton tensor, defined by
\bq
\lb{2.2c}
C^{ij} =  \frac{e^{ikl}}{\sqrt{g}} \nabla_{k}\Big(R^{j}_{l} - \frac{1}{4}R\delta^{j}_{l}\Big),
\eq
with  $e^{123} = 1$, etc. $\lambda, \gamma_n, \beta_{s} $ and $ \Lambda_g$   are the coupling constants of the theory. In terms of $R_{ij}$, we have
\cite{ZSWW},
\bqn
\lb{2.2da}
C_{ij}C^{ij}
&=& \frac{1}{2}R^{3} - \frac{5}{2}RR_{ij}R^{ij} + 3 R^{i}_{j}R^{j}_{k}R^{k}_{i}  +\frac{3}{8}R\Delta R\nb\\
& &  +
\left(\nabla_{i}R_{jk}\right) \left(\nabla^{i}R^{jk}\right) +   \nabla_{k} G^{k},
\eqn
where
\lb{2.2e}
\bqn
G^{k}=\frac{1}{2} R^{jk} \nabla_j R - R_{ij} \nabla^j R^{ik}-\frac{3}{8}R\nabla^k R.
\eqn
%
%
%
 
 It should be noted that in writing the above action, we have excluded all the terms that violate the parity \cite{ZWWS,ZSWW}. For our current   purpose, we  add 
 the fifth and third-order spatial derivative operators to the potential ${\cal{L}}_{V}$ \cite{WWZZ}, 
\bqn \lb{parity action}
\Delta {\cal{L}}_V &=& \frac{1}{M_*^3} \left(\alpha_0 K_{ij} R_{ij} +\alpha_2 \epsilon^{ijk} R_{il} \Delta_j R^l_k \right) \nb\\
&&+ \frac{\alpha_1 \omega_3(\Gamma)}{M_*}+``\dots".
\eqn 
Here the coupling constant $\alpha_0,\;\alpha_1,\;\alpha_2$ are dimensionless and arbitrary, $\epsilon^{ijk}=e^{ijk}/\sqrt{g}$ is the total antisymmetric tensor, and $\omega_3(\Gamma)$  the 3-dimensional gravitational Chern-Simons term \footnote{To take quantum effects into account, it was proposed to add boundary terms $\Delta{S}_{3} = \sum_i{\beta_i M^{3- \Delta_i}\int_{t=t_*} d^3x \sqrt{g} {\cal{O}}^i}$ into the 
Einstein-Hilbert action at the  moment $t = t_*$, right before the inflation started \cite{Porrati}. Clearly, one choice of ${\cal{O}}^i$ is ${\cal{O}}^i \propto \omega(\Gamma)$. We thank Jiro Soda for pointing it
out  to us.}.  ``..." denotes the rest of the fifth-order operators given in Eq.(2.6) of \cite{ZSWW}. Since they have no contributions to tensor perturbations, in this paper we shall not write them out explicitly. 
As shown   in \cite{WWZZ}, because of the additional parity violation terms of Eq.(\ref{parity action}), the non-adiabatic evolution of modes lead to a large polarization of PGWs, and it could be well within the detection of CMB observations, as mentioned above. In this paper, we   investigate their effects on the non-gaussianities of PGWs.

\subsection{The Linearized Tesnor Perturbations}

The general formulas of the linearized tensor perturbations were given in \cite{WWZZ}, so in the rest of this section we give a very brief summary of  the main results obtained there, in order to 
to initiate our studies of the non-gaussianities of PGWs in the next section. 
For details, we refer readers to \cite{WWZZ}.  Consider a flat Friedmann-Robertson-Walker (FRW) universe, 
\bqn
\lb{FRW}
&& \hat{N} = a(\eta),\;\;\; \hat{N}^i = \hat{A} = \hat{\varphi} = 0,\nb\\
&& \hat{g}_{ij}dx^idx^j = a(\eta)^2\delta_{ij}dx^idx^j,
\eqn
where quantities with hats denote the background of the FRW universe in the coordinates $(\eta, x^i) =  (\eta, x, y, z)$. 
Then,   the tensor perturbations are given by,  
\bqn
\lb{pertub}
\delta{N} &=&  \delta{N}^i = \delta{A} = \delta{\varphi} = 0,\nb\\
\delta g_{ij}&=& a^2h_{ij}(\eta, {\bf x}).
\eqn
Assuming that   matter fields have no contributions to tensor perturbations, we find that   the quadratic part of the total action can be cast in the form, 
\bqn
&&S^{(2)}_{\text{ g}}=\zeta^2 \int d\eta d^3 x \Bigg\{\frac{a^2}{4} (h_{ij}')^2-\frac{1}{4} a^2 (\partial_k h_{ij})^2\nb\\
&&\;\;\;\;\;\;\;-\frac{\hat{\gamma_3}}{4 M_*^2}(\partial^2h_{ij})^2-\frac{\hat{\gamma}_5}{4M_*^4 a^2}(\partial^2 \partial_k h_{ij})^2\nb\\
&&\;\;\;\;\;\;\;-\frac{\alpha_1 a e^{ijk}}{2 M_*} (\partial_l h_{i}^m \partial_m \partial_j h_k^l-\partial_l h_{im} \partial^l \partial_j h^m_k)\nb\\
&&\;\;\;\;\;\;\;-\frac{\alpha_2 e^{ijk}}{4M_*^3 a}\partial^2 h_{il} (\partial^2 h^l_k)_{,j} - \frac{3 \alpha_0 {\cal{H}}}{8 M_* a }(\partial_k h_{ij})^2\Bigg\},\nb\\
\eqn
where $h_{ij}' \equiv \partial{h}_{ij}/\partial \eta, \; \partial^2 \equiv \delta^{ij}\partial_{i} \partial_{j},\; 
{\cal{H}} = {a'}/{a}$, and 
$$ 
\gamma_3  \equiv  \left(\frac{M_{pl}}{2M_{*}}\right)^2  \hat{\gamma}_3, \;\;\;
 \gamma_5  \equiv   \left(\frac{M_{pl}}{2M_{*}}\right)^4 \hat{\gamma}_5.
 $$
To avoid fine-tuning,   ${{\alpha}}_{n}$ and  $\hat{\gamma}_{n}$  are expected to be  of  the same order.
Then,   the field equations for $h_{ij}$ read,
 \bqn
 \lb{DFE}
 h''_{ij}&+&2\mathcal{H}h'_{ij}- \alpha^2 \partial^2
 h_{ij}+ \frac{\hat\gamma_3}{a^2M_*^2}\partial^4
 h_{ij}-\frac{\hat\gamma_5}{a^4M_*^4}\partial^6 h_{ij}  \nb\\
&+ &
e_{i}^{\;\; lk}\left(\frac{2\alpha_1}{M_* a} + \frac{\alpha_2}{M_*^3 a^3}\partial^2 \right) \left(\partial^2h_{jk}\right)_{,l} = 0,
 \label{eq7}
 \eqn
where 
$\alpha^2 \equiv 1+ {3\alpha_0{\cal{H}}}/{(2M_*^3 a)}$.

To study the evolution of $h_{ij}$, we expand it    over spatial Fourier harmonics,
\bq
\lb{FT}
 h_{ij}(\eta,{\bf x}) = \sum_{s=R,L} \int \frac{d^3 {\bf
 k}}{(2\pi)^3} \psi_k^s(\eta) e^{i{\bf k}\cdot{\bf x}}P_{ij}^{(s)}(\hat{\bf k}),
\eq
where $P_{ij}^{(s)}(\hat{\bf k})$ are the circular polarization tensors and
satisfy the relations: $i k_m {e}^{rmj} P_{ij}^{(s)}=k \rho^s P_{i}^{r(s)}$ with
$\rho^R=1$, $\rho^{L}=-1$, and ${P^*}_{j}^{i(s)} P_{i}^{j(s')}=\delta^{ss'}$ \cite{soda}.
 Define $u_k^{s} (\eta) = \frac{1}{2}a(\eta) M_{pl} \psi^s_k (\eta)$ and with the de Sitter background $a=-1 / (H \eta)$, we obtain
 \bqn
 \lb{modeu}
 u_k^{s}(\eta)'' +\left[\omega_s^2(k, \eta)- \frac{2}{\eta^2}\right] u_k^{s}(\eta)=0,
 \eqn
 where
 \bqn
 \omega^2_{s}(k, \eta) &\equiv& \alpha^2 k^2 \Big[1-\delta_1 \rho^s (\epsilon_{*} \alpha k\eta )+\delta_2 (\epsilon_{*} \alpha k\eta)^2 \nb\\
 &&+ \delta_3 \rho^s (\epsilon_{*} \alpha k\eta)^3 + \delta_4 (\epsilon_{*} \alpha k \eta)^4 \Big],
 \eqn
 with $\epsilon_{*} \equiv {H}/{M_*} \ll 1$, and 
 \bqn
 \lb{delta1}
 && \delta_1\equiv \frac{2 \alpha_1 }{\alpha^3 },\;\;\;
 \delta_2\equiv \frac{\hat{\gamma_3} }{\alpha^4 },\nb\\
 &&
 \delta_3\equiv \frac{\alpha_2 }{\alpha^5 },\;\;\;
 \delta_4\equiv \frac{\hat{\gamma_5} }{\alpha^6 }.
 \eqn
 
 Following \cite{martin,MB03}, we choose the initial conditions at $\eta=\eta_i$  as
 \bqn
 u_k^{s}(\eta_i)=\frac{1}{\sqrt{2 \omega_{s}(k, \eta_i)}},\;\; u_k^{s}(\eta_i)'=i \sqrt{\frac{\omega_{s}(k, \eta_i)}{2}}.\;
 \eqn
 Then, if one assumes that $\omega_{s}(k,\eta)$ is slowly varying,  i.e., 
 \bqn\lb{wkb}
 {\cal{Q}}\equiv \left|\frac{\omega_{s}(k,\eta)'}{\omega^2_{s}(k,\eta)}\right| \ll 1,
 \eqn
 one can approximatively treat $\omega_{s}(k,\eta)$ as constant, and get the approximate solution of the mode function $u^s_k(\eta)$,
 \bqn
 u_k^{s}(\eta) \simeq  \frac{1}{\sqrt{2 \omega_{s}(k,\eta)}} \left(1- \frac{i}{\omega_{s} \eta }\right) e^{- i  \int \omega_{s} d\eta},
 \eqn
 or,
 \bqn
 \psi_k^{s}(\eta)   \simeq - 2  \frac{ i H}{M_{pl}}\frac{ 1}{\sqrt{2 \omega_{s}^3}} (1+i \omega_{s} \eta) e^{- i \int \omega_{s} d\eta}.
 \eqn
Promoting   $\psi^s_k(\eta)$ to a quantum operator, 
 \bqn
 \psi^{s}_k(\eta) = \psi_k^{s} a_{s}(k)+\psi_{k}^{*{s}}a_{s}^{\dag}(k),
 \eqn
 one finds that
  the power spectrum of the tensor perturbations is given by
 \bqn
\Delta^2_{\text{T}}&\equiv& \frac{k^3 (|\psi^R_k|^2+|\psi^L_k|^2)}{2\pi^2}
\simeq \frac{2 H^2}{\pi^2 M^2_{\text{pl}}},
 \eqn
 which has the same expression as that given  in general relativity. Here $a_{s}(k)$ and $a_{s}^\dag(-k)$ are annihilation and creation operators, and their commutation relation is given by 
 $$
 [a_s(k), a_{s'}^{\dag}(k')]=(2 \pi)^3 \delta_{ss'} \delta(k-k').
 $$
 
 It should be noted that, in our calculations we have assumed (\ref{wkb}). This condition implies that  the adiabatic condition is always satisfied (before the modes exit the horizon), and thus, there is no important modification in the power spectrum of PGWs. Once this condition  is violated, as shown in \cite{WWZZ}, some interesting modifications on power spectrum and polarization of PGWs become possible. For simplification, in this paper  we assume that (\ref{wkb}) always holds.
 
\section{The Interaction Hamiltonian and  bispectrum} 
\renewcommand{\theequation}{3.\arabic{equation}} \setcounter{equation}{0}

In this section, we turn to the cubic action and the bispectrum of the tensor perturbations. The cubic action $S_{\text{g}}^{(3)}$ is given by Eq.(\ref{cubic}), which can be written in the form,
\bqn
S_{\text{g}}^{(3)}=-\int d\eta H_{\text{int}}(\eta).
\eqn
Then the 3-point correlation  function can be computed by employing the in-in formalism \cite{in-in},
\bqn
&&\left<\psi^{s_1}_{k_1}(\eta)\psi^{s_2}_{k_2}(\eta)\psi^{s_3}_{k_3}(\eta)\right>\nb\\
&&\;\;\;=-i \int_{\eta_i}^{\eta} d\eta'\left <[\psi^{s_1}_{k_1}(\eta)\psi^{s_2}_{k_2}(\eta)\psi^{s_3}_{k_3}(\eta), H_{\text{int}}(\eta')]\right>,\nb\\
\eqn
where $\eta_i$ represents the early time when inflation starts, and $\eta$ is a time when the bispectrum is evaluated. A good approximation is to extend the integral into the whole half axis,
$\eta\in(-\infty, 0)$. After some simple but very tedious calculations, it can be shown that the 3-point correlation function can be rewritten in the form
\bqn
&&\left<\psi^{s_1}_{k_1}(0)\psi^{s_2}_{k_2}(0)\psi^{s_3}_{k_3}(0)\right>\nb\\
&&\;\;\;=i (2\pi)^3 \delta^3(k_1+k_2+k_3)\zeta^2 \int_{-\infty}^0 a^2(\eta') d\eta'\nb\\
&&\;\;\;\;\;\times F_{k_1 k_2 k_3}^{s_1 s_2 s_3}(\eta') \Big[W_{k_1 k_2 k_3}^{s_1 s_2 s_3}(\eta')-W_{k_1 k_2 k_3}^{*s_1 s_2 s_3}(\eta')\Big],\nb\\
\eqn
where $F_{k_1 k_2 k_3}^{s_1 s_2 s_3}(\eta')$ is given in Appendix B, and $W_{k_1 k_2 k_3}^{s_1 s_2 s_3}(\eta')$ is defined as
\bqn
 W_{k_1 k_2 k_3}^{s_1 s_2 s_3}(\eta')& \equiv& \psi_{k_1}^{s_1} (0) \psi_{k_2}^{s_2} (0) \psi_{k_3}^{s_3}(0) \nb\\
&&  \times \psi_{k_1}^{*s_1} (\eta') \psi_{k_2}^{*s_2} (\eta') \psi_{k_3}^{*s_3}(\eta').
\eqn
In the   de Sitter background,   the 3-point correlation function reduces to, 
\bqn
&&\left<\psi^{s_1}_{k_1}(0)\psi^{s_2}_{k_2}(0)\psi^{s_3}_{k_3}(0)\right>\nb\\
&&\;\;\;= (2\pi)^7 \delta^3(k_1+k_2+k_3) \frac{\Delta^4_{\text{T}}}{2^3k_1^3 k_2^3 k_3^3} B^{s_1 s_2 s_3 }_{k_1,k_2,k_3},
\eqn
where 
\bqn\lb{bs}
&&B^{s_1 s_2 s_3}_{k_1 k_2 k_3}\equiv \sum_{n=0}^{4} \delta_n \epsilon_{*}^n  F_n I_{n},
\eqn
with $\delta_0=1$,  and $I_{n}$ is given by, 
\bqn\lb{mode}
I_{n}&\equiv& \text{Im} \Bigg\{\int_{-\infty}^0 d\eta (-\eta)^{n-2} e^{i \int (\omega_{s_1}+\omega_{s_2}+\omega_{s_3})d\eta}\nb\\
&&\;\;\;\;\;\;\;\;\;\;\times \sqrt{\frac{k_1^3 k_2^3 k_3^3}{\omega_{s_1}^3\omega_{s_2}^3\omega_{s_3}^3}} (1-i\omega_{s_1}\eta)\nb\\
&&\;\;\;\;\;\;\;\;\;\times (1-i\omega_{s_2}\eta)(1-i\omega_{s_3}\eta)\Bigg\}.
\eqn

Usually, the $k$-dependence of the bispectrum receives contributions from both the interaction Hamiltonian $H_{\text{int}} \left(\sim \sum F_n\right) $ and the mode function integration $I_n$.  For the former, one can see from (\ref{bs}) that the high order spatial derivative terms do have contributions in bispectrum, but   are suppressed by the factor $\epsilon_{*}$. Then, the leading-order contributions come from the two derivative term $F_0$, which has the same expression as that given  in general relativity. 

On the other hand, the mode function integration $I_n$ in the current case involved very complicated expression, and thus it is very hard to perform the integration explicitly. In general relativity, to minimize the errors, one  usually splits the integrals into three different regions: one  outside the horizon, one around the horizon, and one  inside the horizon. The contribution from the last region vanishes due to the high-frequency oscillate \cite{maldacena}. A similar conclusion is also applicable to  the current case,  because when the high-order derivative terms dominate,  the oscillation becomes more rapidly than that in general relativity. Thus, using the same arguments,  one can safely neglect the effects from the high-order derivative terms in the last region. 

In the first region, the mode is outside the horizon and the effective mass term $-{2}/{\eta^2}$ in Eq.(\ref{modeu})   dominates. Then, in this region the corresponding results  are also the same as those given  in general relativity. Therefore, the effects from the parity-violating operators  come only from the region around the horizon. In this region, although the $k^2$ term dominates, the high-derivative terms still have non-negligible contributions. In order to take these effects into account, 
we expand the integration in terms of $\epsilon_{*}$.  In particular, to the zeroth-order of $\epsilon_*$, 
we have
\bqn
&&\text{Im}\Bigg\{i \int_{-\infty}^0 d\eta (-\eta)^{n-2} e^{i(k_1+k_2+k_3)\eta} \nb\\
&&\;\;\;\times (1+i k_1 \eta)(1+i k_2 \eta)(1+i k_3 \eta)\Bigg\},
\eqn
which coincides with the mode integration in general relativity, as expected. Thus, one immediately
obtains the bispectrum of the leading-order,
\bqn
&&B_{\text{(GR)}}^{s_1 s_2 s_3}(k_1, k_2, k_3) = \nb\\
&&\;\;\;\; \left(-K+\frac{k_1 k_2+k_1 k_3+k_2 k_3}{K}+\frac{k_1 k_2 k_3}{K}\right) F_0,\nb\\
\eqn
where $K\equiv k_1+k_2+k_3$, and which is precisely  the bispectrum of PGWs given in GR.

Now let us turn to the first order contributions  of $ \epsilon_{*}$. Ignoring all the detailed   calculations,  it can be shown that it takes the form,  
\bqn\lb{PVcontribution}
&&B_{\text{(PV)}}^{s_1 s_2 s_3}(k_1, k_2, k_3) = \nb\\
&&\;\;-\frac{\pi}{2} \delta_1 \epsilon_{*} \left[F_1+\frac{3}{4} (s_1 k_1+s_2k_2+s_3 k_3) F_0\right]. ~~~~
\eqn
Since this term is directly proportional to $\delta_1$, from Eqs.(\ref{parity action}) and (\ref{delta1}) we find that it represents  the contributions of the three-dimensional Chern-Simons term. 

For  operators with dimensions  $n \ge 4$, their contributions can be written in the form,  
\bqn\lb{int}
\epsilon_*^{n-2} \sum_{r}{F_r \text{Im}\Bigg\{\int_{-\infty}^0 d \eta (-\eta)^{n-4} e^{iK \eta}f_{n -r -2}(\eta)\Bigg\}},\nb\\
\eqn
where $ r = (0, n -2)$,  and $f_{n-r-2} (K,\eta)$   can be expressed as
\bqn
f_{n -r -2} (K,\eta)&\equiv& a_0  +a_1  (i \eta)+a_2(i\eta)^2+\dots \nb\\
&&+ a_{n-r+1} (i\eta)^{n-r+1},
\eqn
here $a_r$ are functions of $k_1,\;k_2,\;k_3$.   In particular, the  effects from fifth-order derivative terms should contribute to the bispectrum at the order of $\varepsilon^3_{*}$. But, a careful analysis over the integration shows that   their contributions      vanish identically.  This result can be easily generalized to higher order terms. In fact,  for $n = 2j + 1$ with $j  = 2, 3, 4, ...$,  the  bispectrum of PGWs  at the order $\epsilon_{*}^{n-2}$ always vanishes. This implies that $B_{\text{(PV)}}^{s_1 s_2 s_3}(k_1, k_2, k_3)$ given by Eq.(\ref{PVcontribution}) represents the only contribution from parity violation operators. 

In ref.\cite{parity-soda}, Soda {\em et al.} calculated the bispectrum of PGWs from the Weyl cubic terms, $W^3$ and $\tilde{W}W^2$, and proved that  no contributions from parity violation appear in the non-gaussianity of PGWs in pure de Sitter background.  This is consistent with our current results. However, it must be noted that   in their considerations, the symmetry of the theory is still of the general  diffeomorphisms. As a result,   only   the parity-violating terms  $W^3$ and $\tilde{W}W^2$  are allowed. These  terms   are both P-odd and T-odd. Thus, when one calculates the bispectrum, the two terms produce an integral similar to (\ref{int}) with $n-2 = 2j -1$ as odd number that is greater than two \cite{parity-soda,parity2}. Hence,  with the arguments given above, they indeed have no contributions to the bispectrum of PGWs.

 \begin{figure}[!t]
\centering
	\subfigure[]
	{\label{GRb1}\includegraphics[width=65mm]{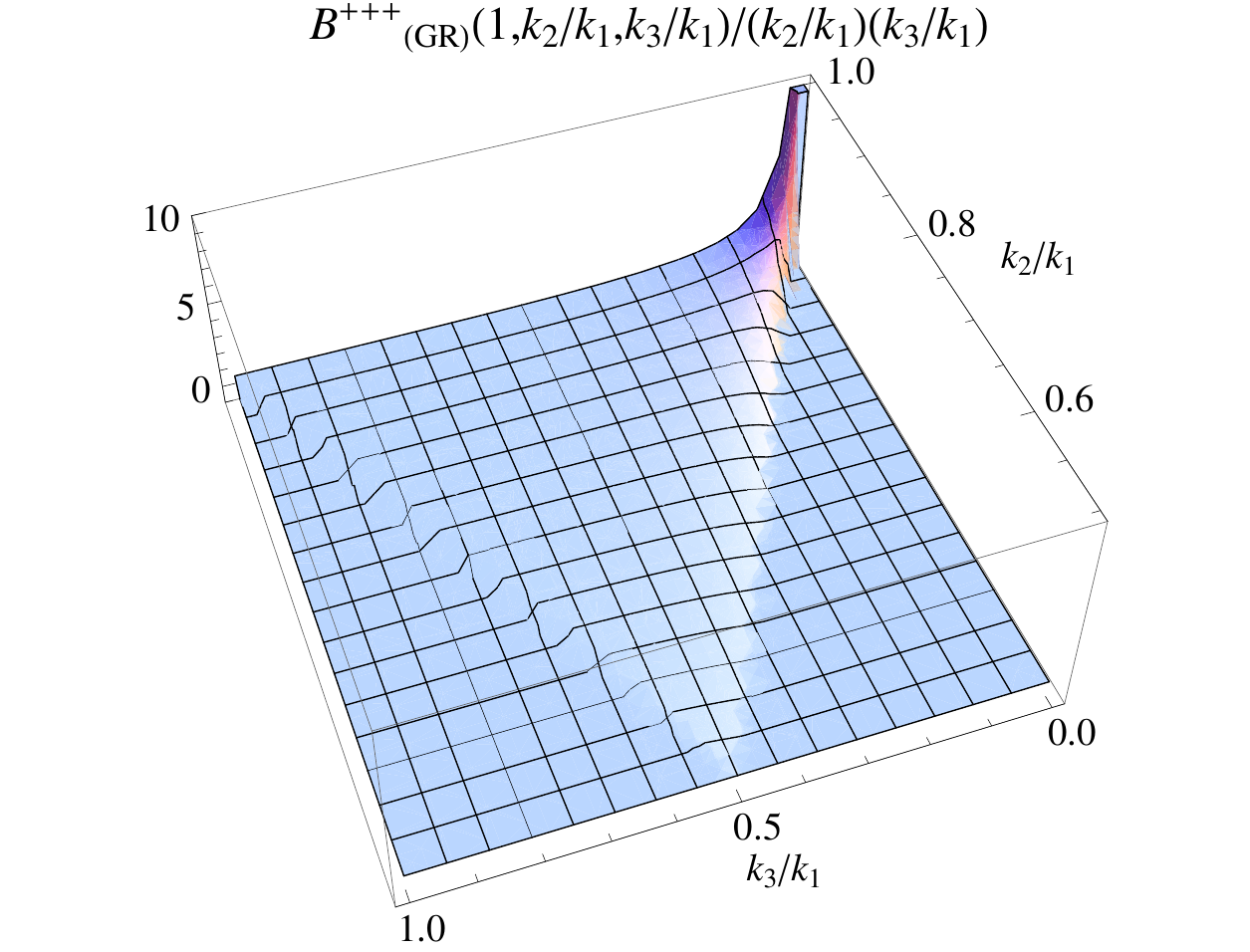}}\\
	\subfigure[]
	{\label{GRb2}\includegraphics[width=65mm]{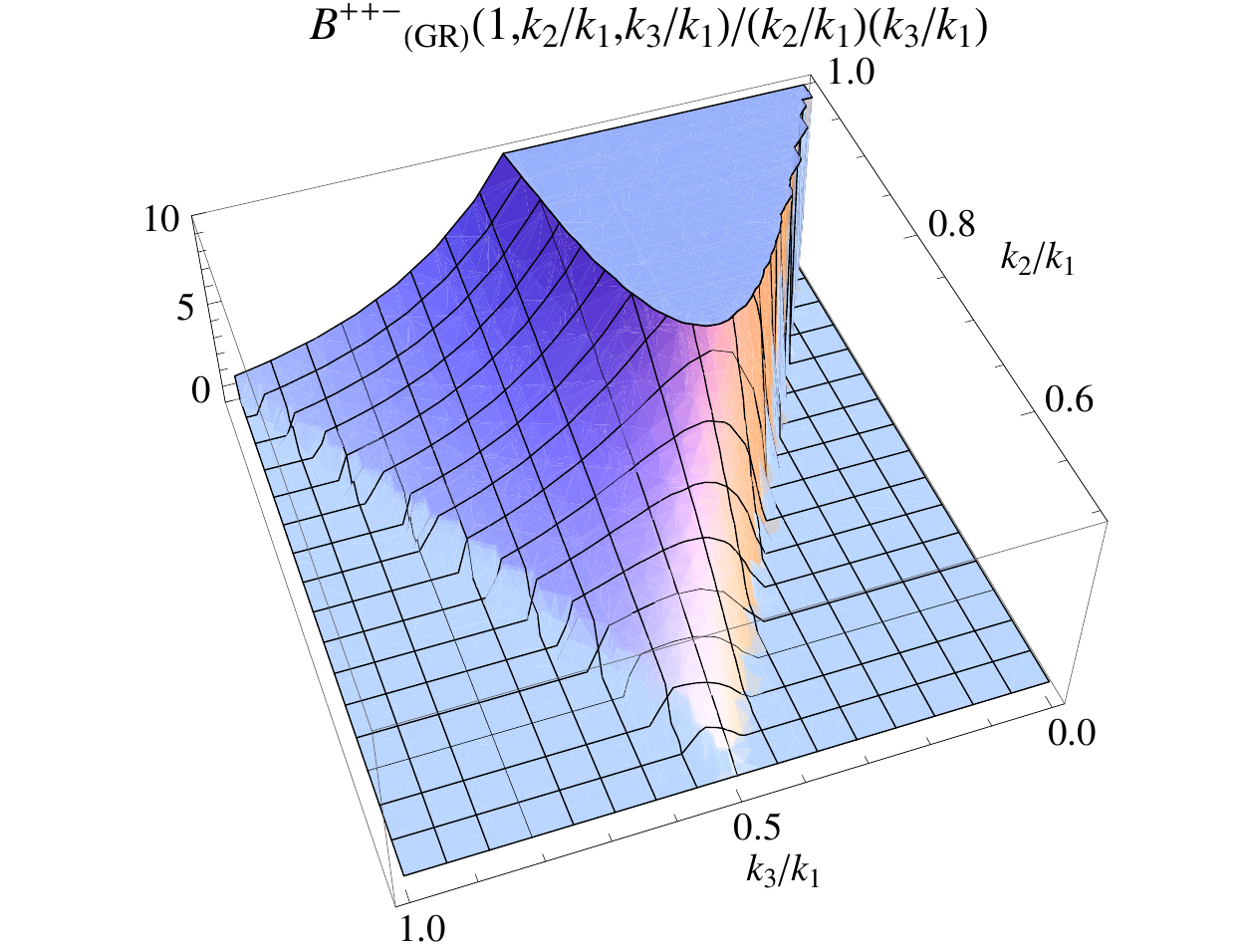}}
\caption{(a) Shape of $(k_1k_2k_3)^{-1}B_{\text{(GR)}}^{+++} (k_1,k_2,k_3)$. (b) Shape of $(k_1k_2k_3)^{-1}B_{\text{(GR)}}^{++ -} (k_1,k_2,k_3)$.  All are normalized to unity in the equilateral limit.}
\lb{GR}
\end{figure}

\section{Shape of the Bispectrum}
\renewcommand{\theequation}{4.\arabic{equation}} \setcounter{equation}{0}

We are now ready to plot the shapes of the bispectrum. For $s_1 = s_2 = s_3 = 1$ and $s_1 = s_2 = -s_3 = 1$, we plot the shapes of the bispectrum of the leading order conrtibutions $(k_1 k_2 k_3 )^{-1}B^{s_1 s_2 s_3}_{\text{(GR)}}(k_1,k_2,k_3)$ in Fig. \ref{GR}. Because there is no parity violation in the leading order, we have $B^{+++}_{\text{(GR)}}(k_1,k_2,k_3)=B^{---}_{\text{(GR)}}(k_1,k_2,k_3)$ and $B^{++-}_{\text{(GR)}}(k_1,k_2,k_3)=B^{--+}_{\text{(GR)}}(k_1,k_2,k_3)$. Thus, there are only two   possible configurations, which both peak at the squeezed limit ($k_3/k_1\rightarrow 0$). As pointed out in \cite{Gao-PRL}, the second configuration (Fig.(\ref{GRb2})) is sub-dominant, in comparison   with the  first one in the equilateral limit ($k_1\simeq k_2 \simeq k_3 $), that is, $B^{++-}_{\text{(GR)}}\simeq B^{+++}_{\text{(GR)}}/81$.

Now we turn to the contributions from the parity-violating Chern-Simons term. In this case, because of the violation of the parity, for different spin products we have four independent configurations. We plot the shapes of these four configurations in Figs.(\ref{PV}), from which it can be seen  that all the configurations peak in the squeezed limit (Note that for the $(++-)$ and $(---)$ cases they peak in the negative direction).  More specifically, we have $B_{\text{(PV)}}^{+++}(k_1, k_2 ,k_3)=-B_{\text{(PV)}}^{---}(k_1, k_2 ,k_3)$, and $B_{\text{PV}}^{++-}(k_1, k_2 ,k_3)=-B_{\text{(PV)}}^{--+}(k_1, k_2 ,k_3)$.

\begin{figure}[!t]
\centering
	\subfigure[]
	{\label{PVa}\includegraphics[height=45mm]{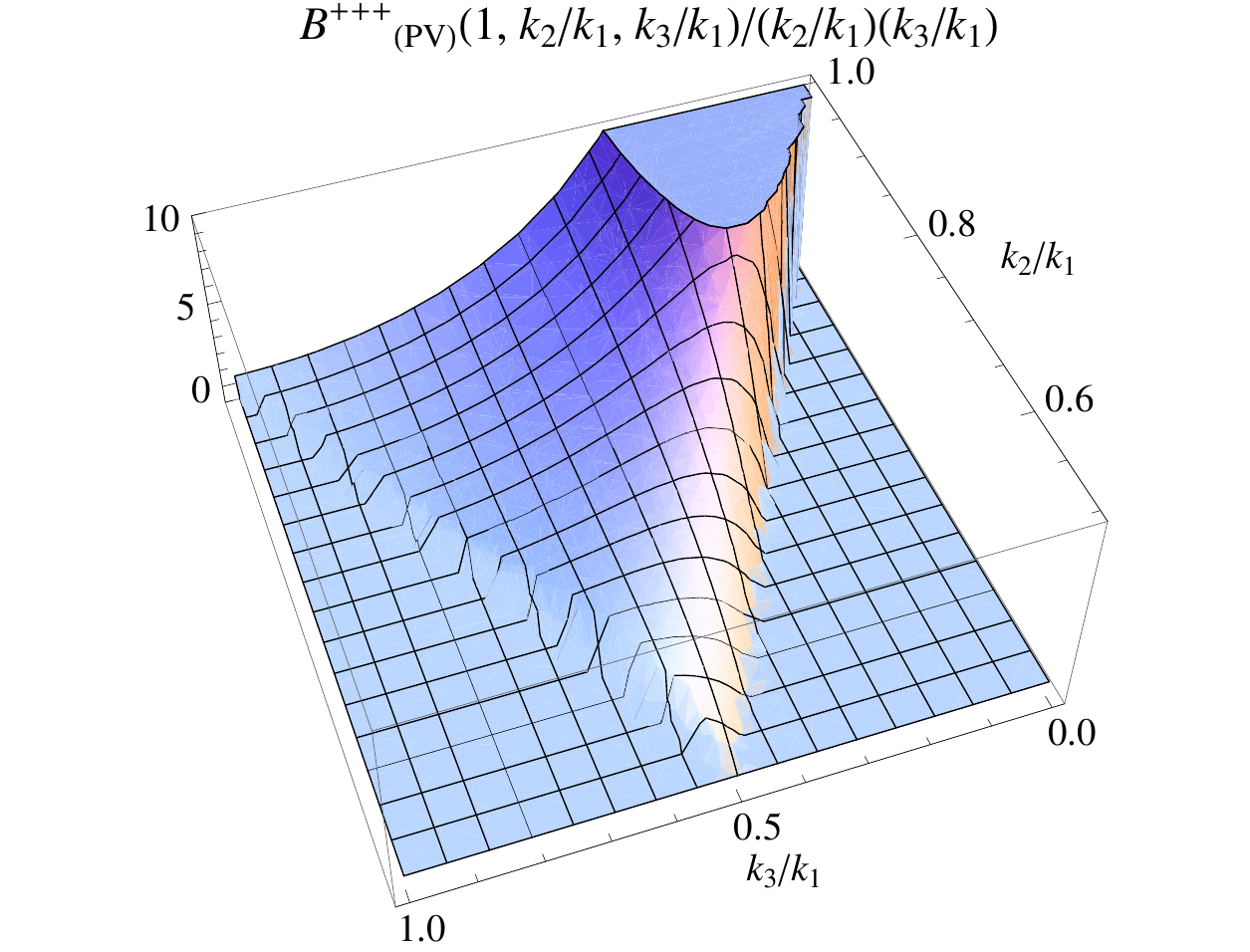}}\\
	\subfigure[]
	{\label{PVb}\includegraphics[height=45mm]{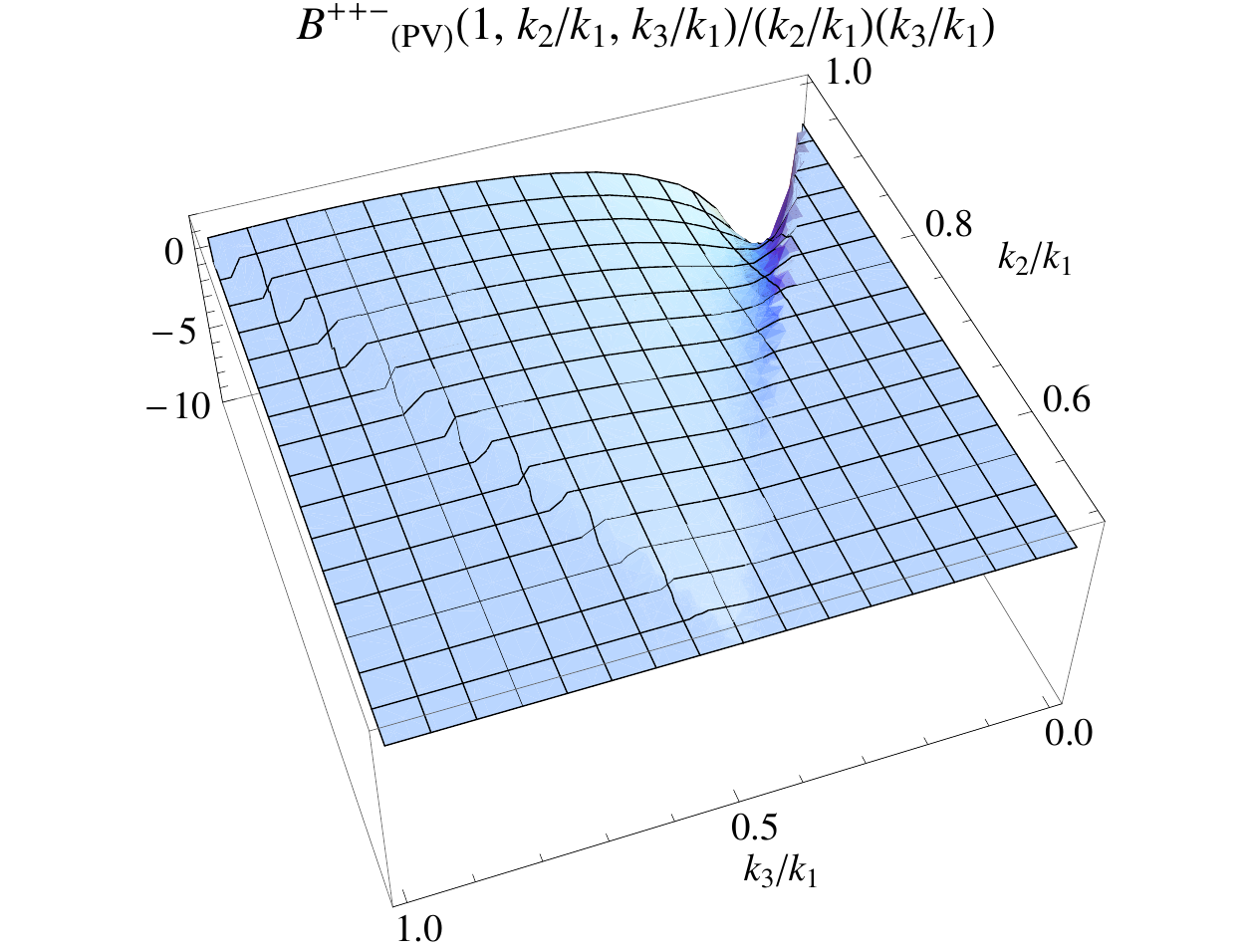}}\\
	\subfigure[]
	{\label{PVc}\includegraphics[height=45mm]{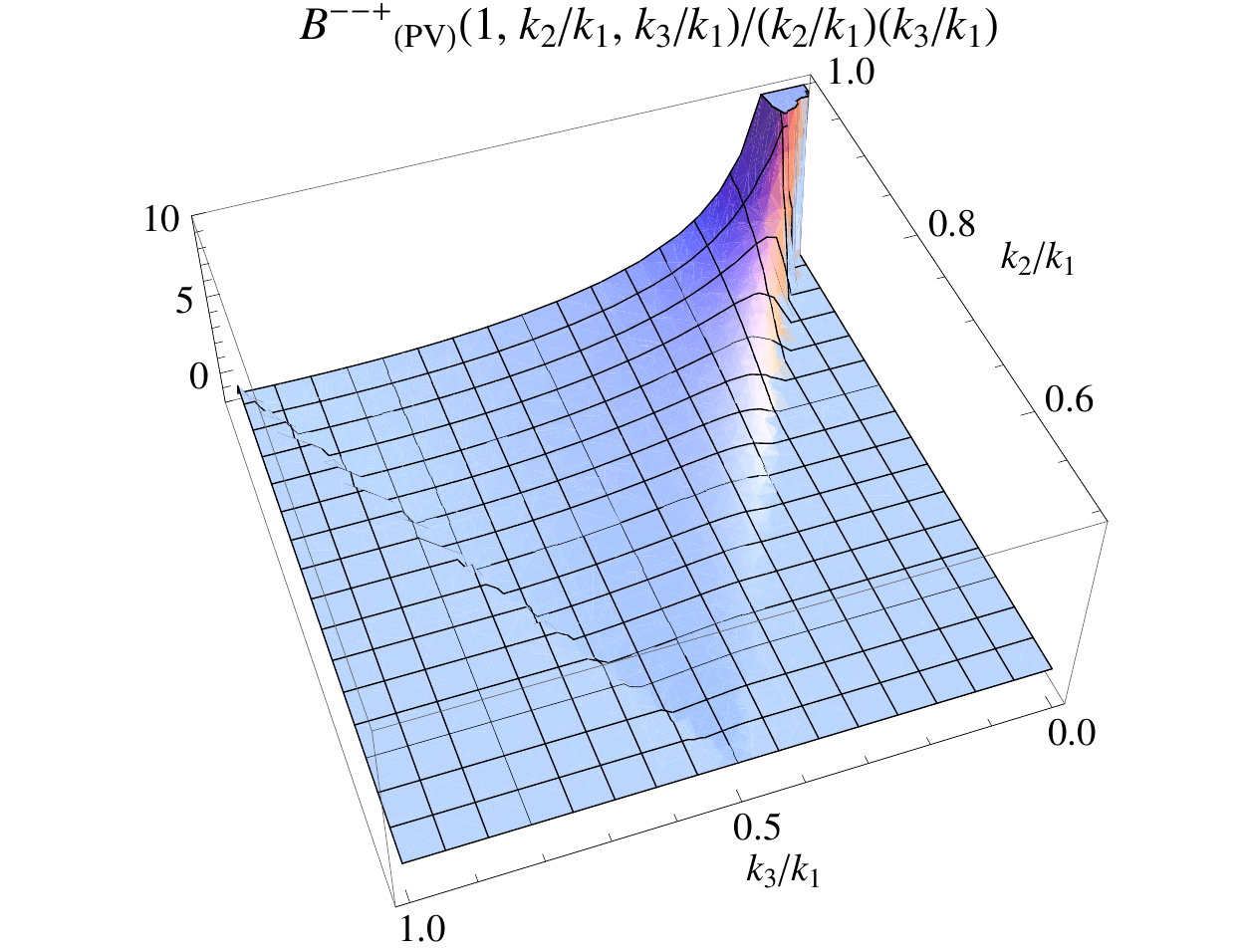}}\\
	\subfigure[]
	{\label{PVd}\includegraphics[height=45mm]{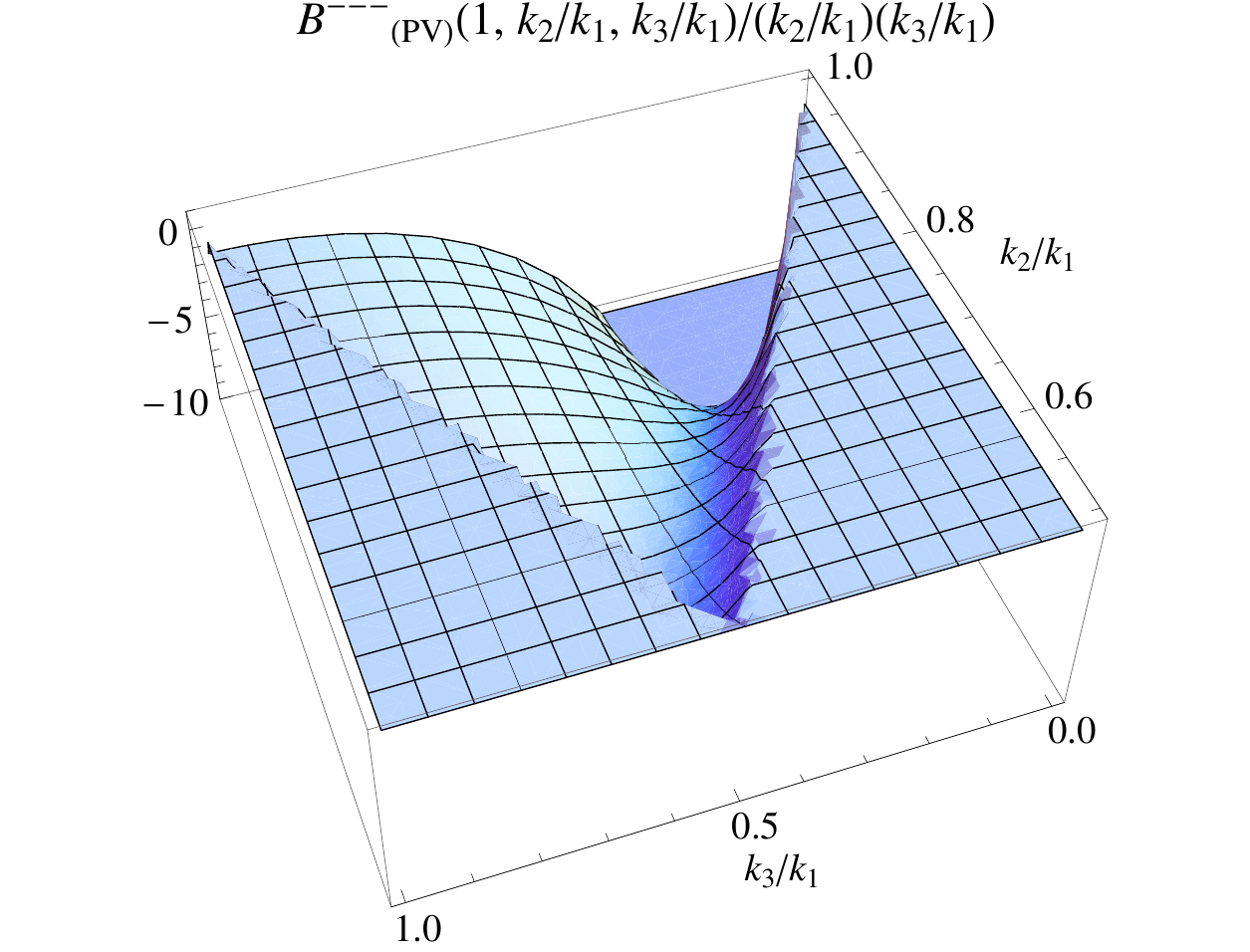}}
\caption{Shapes of $ (k_1 k_2 k_3 )^{-1}B^{s_1 s_2 s_3}_{\text{(PV)}}(k_1,k_2,k_3)$ for various spin products: (a) $+++$; (b) $++-$; (c) $--+$; (d) $---$. 
All are normalized to unity in the equilateral limit.}
\lb{PV}
\end{figure}

\section{Conclusions and Remarks}
\renewcommand{\theequation}{5.\arabic{equation}} \setcounter{equation}{0}

In this paper, we have investigated the non-gaussianities of PGWs generated during the de Sitter expansion of the universe in the framework of the HL theory, and paied particular attention on the effects of the operators that violate the parity. Because of the restricted foliation-preserving diffeomorphisms  of the theory, the parity-violating third and fifth-dimensional  operators exist generically.   By calculating the three-point correlation  function of the PGWs, we have shown that the leading-order contributions still  come from the second-order spatial derivative terms, and are  the same as those given in general relativity.  The high-order n-th spatial derivative terms of the theory also contribute to the bispectrum, although their contributions are suppressed by an factor $\epsilon_{*}^{n-2}$. 

More remarkably, we have also found that  the three-dimensional gravitational Chern-Simons operator  $\omega(\Gamma)$ is the only one that violates the parity and meantime has non-vanishing contributions to the  non-gayssianities of PGWs. In comparison with the contributions of the second-order operators that produce the same non-gaussianity as given in general relativity, its contributions are suppressed by the factor $ \epsilon_{*}$.  In addition, operators with odd-order and higher than three have no contributions to  the non-gaussianities of PGWs.

It should be noted that in obtaining the above results, we have assumed that the adiabatic condition (\ref{wkb}) is always satisfied before the modes exit the horizon. If this condition fails to hold, the non-adiabatic evolution of the modes becomes possible, and hence the integral history of the mode function will be dramatically altered. Then, large non-gaussianities are expected, although it is still an open question how  one can extract information of PGWs in this case. 

\section*{Acknowlodgements}

We would like to express our gratitude to Shinji Mukohyama and Jiro Soda for valuable discussions and comments.  This work was supported in part by DOE, DE-FG02-10ER41692 (A.W.);  
 the  NSFC with   Grants No. 11205133 (Q.W. and T.Z.), No. 11047008 (Q.W. and T.Z.), No. 11105120 (T.Z.),   No. 11173021 (W.Z. $\&$ A.W.), No. 11075141 (W.Z. $\&$ A.W.),  No. 11375153 (A.W.), No. 11322324 (W.Z.);  Ci\^encia Sem Fronteiras, No. 004/2013 - DRI/CAPES (A.W.);   and the project of Knowledge Innovation Program of Chinese Academy of Science (W.Z.).

\section*{Appendix A:  The Cubic Action of PGWs}
\renewcommand{\theequation}{A.\arabic{equation}} \setcounter{equation}{0}

The cubic action for the tensor perturbations can be  written in the form, 
\bqn\lb{cubic}
S^{(3)}_{\text{g}} &=& \zeta^2 \int d\eta d^3 x  a^2 \Bigg\{\frac{\alpha^2}{4}L_2+\frac{\alpha_1}{a M_*} L_3+\frac{\hat{\gamma}_3 }{a^2 M_*^2}L_4\nb\\
&&\;\;\;\;+\frac{\alpha_2 }{a^3 M_*^3}L_5+\frac{\hat{\gamma}_5 }{a^4 M_*^4}L_6\Bigg\},
\eqn
where
\bqn
L_2&=& h^{mk, ij} (2 h_{mi} h_{kj} - h_{ij} h_{mk} ) ,\nb\\
L_3&=& - \frac{1}{2} e^{ijk} (\partial^2 h_{lk})_{,j} h_{im} h^{lm}\nb\\
&& - \frac{e^{ijk}}{4}  \Big[4 h_{i[m,l]j} h^{lp} (h^m_{p,k}+h^{,m}_{pk}-h^m_{k,p}) \nb\\
&&~~~~~~~~~ + h^n_{i,l} h^l_{j,m} \left(\frac{2}{3}h^m_{k,n}-2h_{kn}^{,m}\right) \nb\\
&&~~~~~~~~~ +2 h^n_{l,i} h^l_{m,j} h^m_{k,n} \Big]\nb\\
L_4&=&- \frac{1}{2} (\partial^2 h_{ij})_{,n}^{\;m} h^i_m h^j_n\nb\\
&& - \frac{1}{2}(\partial^2 h_{ij}) h_{mn} \left(2 h^{im,jn} -\frac{h^{mn,ij}}{2}-h^{ij,mn}\right),\nb\\
L_5&=&\frac{e^{ijk}}{2} \Bigg[ -(\partial^2 h_{il} )h^{mp}_{,j} \left( h^l_{p,mk} - \frac{1}{2}h_{mp,k}^{\;\;\;\;\;l} \right)\nb\\
&&~~~~~~~~~ + (\partial^2 h_{il,j}) h^{mp} \left(h_{mk,p}^{\;\;\;\;\;\;l} - h^l_{k,mp}\right) \nb\\
&&~~~~~~~~~ + 2 (\partial^2 h_{il,j}) h_{mk,p}  h^{l[p,m]} \nb\\
&&~~~~~~~~~-\frac{1}{2} (\partial^2 h_{il,j}) h^{lp} \partial^2 h_{pk}\Bigg]\nb\\
&& - \frac{1}{8} e^{ijk} (\partial^2 h_{il}) (\partial^2 h^m_k) (h_{m,j}^l-h^l_{j,m}-h_{jm}^{\;\;\;,l})\nb\\
&& - \frac{1}{4} e^{ijk} (\partial^4 h_{kl,j}) h_{im} h^{lm},\nb\\
L_6&=&\frac{3}{8} (\partial^2 h^i_j) (\partial^2 h^j_k) (\partial^2 h^k_i)\nb\\
&& +\frac{1}{2} \Bigg[(\partial^4 h_{ij}) h_{mn} \Big(2 h^{im,jn} -\frac{1}{2} h^{mn,ij}-h^{ij,mn}\Big)\nb\\
&&~~~~~ +(\partial^4 h_{ij,n}^{\;\;\;m})h^i_m h^{jn} +2 (\partial^2 h_{jk})_{,il} (\partial^2h^{l k}) h^{ij} \Bigg]. \nb\\
\eqn
Then the interaction Hamiltonian is
\bqn
H_{\text{int}}(\eta)&=&\int d^3x {\cal{H}}_{\text{int}}(\eta,x), \nb\\
{\cal{H}}_{\text{int}}(\eta,x)&=&-\zeta^2 a^2 \Bigg[\frac{\alpha^2}{4}L_2+\frac{\alpha_1}{a M_*} L_3+\frac{\hat{\gamma}_3 }{a^2 M_*^2}L_4\nb\\
&&~~~~~~~~+\frac{\alpha_2 }{a^3 M_*^3}L_5+\frac{\hat{\gamma}_5 }{a^4 M_*^4}L_6\Bigg].
\eqn

\section*{Appendix B:  Expression of $F_{k_1k_2k_3}^{s_1 s_2 s_3}(\eta)$}
\renewcommand{\theequation}{B.\arabic{equation}} \setcounter{equation}{0}

$F_{k_1k_2k_3}^{s_1 s_2 s_3} (\eta)$ is given by,
\bqn
F_{k_1k_2k_3}^{s_1 s_2 s_3} (\eta)&=& \alpha^2 F_0 + \frac{\alpha_1}{a M_*} F_1+\frac{\hat{\gamma}_3 }{a^2 M_*^2}F_2\nb\\
&&+\frac{\alpha_2 }{a^3 M_*^3}F_3+\frac{\hat{\gamma}_5 }{a^4 M_*^4}F_4\nb\\
&=& \alpha^2 \sum_{n=0}^{4} \delta_n  \epsilon_{*}^n (-\eta)^n F_{n},
\eqn
where the even parity terms are
\bqn
F_0 &=& {\mathfrak{F}}_k (k_1 s_1+k_2 s_2+k_3 s_3)^4, \nb\\
F_2 &=& {\mathfrak{F}}_k (k_1 s_1+k_2 s_2+k_3 s_3)^4 \alpha^2 \left(k_1^2+k_2^2+k_3^2\right), \nb\\
\eqn
\begin{widetext}
\bqn
F_4&=& \alpha^4 {\mathfrak{F}}_k \Bigg\{k_1^8+4 k_1^7 s_1 (k_2 s_2+k_3 s_3) + 6 k_1^6 (k_2 s_2+k_3 s_3)^2\nb\\
&&~~~~~~~~ +4 k_1^5 s_1 \left(k_2^3 s_2+4 k_2 k_3^2 s_2+4 k_2^2 k_3 s_3+k_3^3 s_3\right)\nb\\
&&~~~~~~~~ +k_1^4 \left(2 k_2^4+19 k_2^2 k_3^2+2 k_3^4 +16 k_2 k_3s_2 s_3 \left(k_2^2+k_3^2\right) \right)\nb\\
&&~~~~~~~~ +2 k_1^3 s_1 \left(2 k_2^5 s_2+8 k_2^4 k_3 s_3+9 k_2^3 k_3^2 s_2+9 k_2^2 k_3^3 s_3+8 k_2 k_3^4 s_2+2 k_3^5 s_3\right) \nb\\
&&~~~~~~~~+k_1^2 \Big[6 k_2^6+19 k_2^4 k_3^2+19 k_2^2 k_3^4+6 k_3^6 +2 k_3s_3s_2 \left(8 k_2^5+9 k_2^3 k_3^2+8 k_2 k_3^4\right)\Big]\nb\\
&&~~~~~~~~ +4 k_1s_1 \left(k_2^4+k_2^2 k_3^2+k_3^4\right)(k_2 s_2+k_3 s_3)^3\nb\\
&&~~~~~~~~ +\left(k_2^4+k_3^4\right) \left(k_2s_2 + k_3s_3\right)^4\Bigg\},\\
{\mathfrak{F}}_k &\equiv& -\frac{(k_1-k_2-k_3)(k_1+k_2-k_3) (k_1-k_2+k_3)(k_1+k_2+k_3)}{256 k_1^2 k_2^2 k_3^2},
\eqn
and the odd parity terms are
\bqn
F_1&=& - \alpha {\mathfrak{F}}_k\Big\{k_1^5 (4 s_1-4 s_2-5 s_3) + k_2^5 (4 s_2-4 s_3-5s_1) + k_3^5 (4 s_3 - 4s_1 - 5s_2)\nb\\
&&~~~~~~~~ -2 k_1^3 k_2^2 (s_1-5 s_2-s_3+2 s_1 s_2 s_3) -2 k_1^3 k_3^2 (s_1 - 4 s_3 + 2 s_1 s_2 s_3)\nb\\
&&~~~~~~~~ -2 k_2^3 k_3^2 (s_2-5 s_3-s_1+2 s_1 s_2 s_3) -2 k_2^3 k_1^2 (s_2 - 4 s_1 + 2 s_1 s_2 s_3)\nb\\
&&~~~~~~~~ -2 k_3^3 k_1^2 (s_3-5 s_1-s_2+2 s_1 s_2 s_3) -2 k_3^3 k_2^2 (s_3 - 4 s_2 + 2 s_1 s_2 s_3)\nb\\
 &&~~~~~~~~ +k_1^4 k_2 (-3 s_1 + 6 s_2 - 4 s_3 - 4 s_1 s_2 s_3) +k_1^4 k_3 (-6 s_1 - 5 s_2 + 6 s_3 - 4 s_1 s_2 s_3)\nb\\
 &&~~~~~~~~ +k_2^4 k_3 (-3 s_2 + 6 s_3 - 4 s_1 - 4 s_1 s_2 s_3) +k_2^4 k_1 (-6 s_2 - 5 s_3 + 6 s_1 - 4 s_1 s_2 s_3)\nb\\
 &&~~~~~~~~ +k_3^4 k_1 (-3 s_3 + 6 s_1 - 4 s_2 - 4 s_1 s_2 s_3) +k_3^4 k_2 (-6 s_3 - 5 s_1 + 6 s_2 - 4 s_1 s_2 s_3)\nb\\
 &&~~~~~~~~ +2 k_1^2 k_2^2 k_3 (5 s_1 + 4 s_2 - 2 s_3 + 18 s_1 s_2 s_3) +2 k_1^3 k_2 k_3 (-7 s_1 + 7 s_2 + 7 s_3 + 2 s_1  s_2 s_3)\nb\\
 &&~~~~~~~~ +2 k_2^2 k_3^2 k_1 (5 s_2 + 4 s_3 - 2 s_1 + 18 s_1 s_2 s_3) +2 k_2^3 k_3 k_1 (-7 s_2 + 7 s_3 + 7 s_1+ 2 s_1 s_2 s_3)\nb\\
 &&~~~~~~~~ +2 k_3^2 k_1^2 k_2 (5 s_3 + 4 s_1 - 2 s_2 + 18 s_1 s_2 s_3) +2 k_3^3 k_1 k_2 (-7 s_3 + 7 s_1 + 7 s_2 + 2 s_1 s_2 s_3)
\Big\},
\eqn
\bqn
F_3&=&\frac{\alpha^3 {\mathfrak{F}}_k}{2 k_1^2 k_2^2 k_3^2 } \nb\\&& \Big\{ k^{11}_1 k^2_3 [2s_2] + k^9_1 k_2 k^3_3 [-8s_3] + k^9_1 k^2_2 k^2_3 [5s_1-8s_2-2s_3] + k^9_1 k^4_3 [-4s_2] \nb\\
&& + k^{11}_2 k^2_1 [2s_3] + k^9_2 k_3 k^3_1 [-8s_1] + k^9_2 k^2_3 k^2_1 [5s_2-8s_3-2s_1] + k^9_2 k^4_1 [-4s_3] \nb\\&& + k^{11}_3 k^2_2 [2s_1] + k^9_3 k_1 k^3_2 [-8s_2] + k^9_3 k^2_1 k^2_2 [5s_3-8s_1-2s_2] + k^9_3 k^4_2 [-4s_1] \nb\\
&& + k^8_1 k^2_2 k^3_3 \left[ 8s_1s_3 (s_1 - s_3) \right] + k^8_1 k^3_2 k^2_3 \left[ 9 + 8s_1 (-s_2 + s_3) \right] + k^7_1 k_2 k^5_3 [8s_3] + k^7_1 k^6_3 [2s_2] \nb\\
&& + k^8_2 k^2_3 k^3_1 \left[ 8s_2s_1 (s_2 - s_1) \right] + k^8_2 k^3_3 k^2_1 \left[ 9 + 8s_2 (-s_3 + s_1) \right] + k^7_2 k_3 k^5_1 [8s_1] + k^7_2 k^6_1 [2s_3] \nb\\
&& + k^8_3 k^2_1 k^3_2 \left[ 8s_3s_2 (s_3 - s_2) \right] + k^8_3 k^3_1 k^2_2 \left[ 9 + 8s_3 (-s_1 + s_2) \right]  + k^7_3 k_1 k^5_2 [8s_2] + k^7_3 k^6_2 [2s_1] \nb\\&& + k^7_1 k^2_2 k^4_3 [-5s_1+20s_2-4s_3] + k^7_1 k^4_2 k^2_3 [2(-3s_1+s_2+10s_3)] + k^7_1 k^3_2 k^3_3 \left[ 4s_2s_3 (3s_1+4s_2+2s_3) \right] \nb\\&& + k^7_2 k^2_3 k^4_1 [-5s_2+20s_3-4s_1] + k^7_2 k^4_3 k^2_1 [2(-3s_2+s_3+10s_1)] + k^7_2 k^3_3 k^3_1 \left[ 4s_3s_1 (3s_2+4s_3+2s_1) \right]  \nb\\&& + k^7_3 k^2_1 k^4_2 [-5s_3+20s_1-4s_2] + k^7_3 k^4_1 k^2_2 [2(-3s_3+s_1+10s_2)] + k^7_3 k^3_1 k^3_2 \left[ 4s_1s_2 (3s_3+4s_1+2s_2) \right]  \nb\\%
&& + k^6_1 k^2_2 k^5_3 [12s_1-2s_2-23s_3] + k^6_1 k^5_2 k^2_3 [14s_1-25s_2-4s_3+8s_1s_2s_3] \nb\\&& + k^6_2 k^2_3 k^5_1 [12s_2-2s_3-23s_1] + k^6_2 k^5_3 k^2_1 [14s_2-25s_3-4s_1+8s_1s_2s_3] \nb\\&& + k^6_3 k^2_1 k^5_2 [12s_3-2s_1-23s_2] + k^6_3 k^5_1 k^2_2 [14s_3-25s_1-4s_2+8s_1s_2s_3] \nb\\%
&& + k^6_1 k^3_2 k^4_3 [(2s_2)(1+4s_1(s_2+s_3))] + k^6_1 k^4_2 k^3_3 [s_3(-3+8s_1(2s_2+s_3))] + k^5_1 k^5_2 k^3_3 [-12s_1+8s_2+16s_3-8s_1s_2s_3] \nb\\&& + k^6_2 k^3_3 k^4_1 [(2s_3)(1+4s_2(s_3+s_1))] + k^6_2 k^4_3 k^3_1 [s_1(-3+8s_2(2s_3+s_1))] + k^5_2 k^5_3 k^3_1 [-12s_2+8s_3+16s_1-8s_1s_2s_3] \nb\\&& + k^6_3 k^3_1 k^4_2 [(2s_1)(1+4s_3(s_1+s_2))] + k^6_3 k^4_1 k^3_2 [s_2(-3+8s_3(2s_1+s_2))] + k^5_3 k^5_1 k^3_2 [-12s_3+8s_1+16s_2-8s_1s_2s_3] \nb\\
&&~ + k^5_1 k^5_2 k^3_3 [(8s_3)(2-s_1s_2-s_1s_3-s_2s_3)] + k^5_1 k^3_2 k^5_3 [4(4s_1-s_3)] \nb\\%
&&~ + k^5_2 k^5_3 k^3_1 [(8s_1)(2-s_1s_2-s_1s_3-s_2s_3)] + k^5_2 k^3_3 k^5_1 [4(4s_2-s_1)] \nb\\%
&&~ + k^5_3 k^5_1 k^3_2 [(8s_2)(2-s_1s_2-s_1s_3-s_2s_3)] + k^5_3 k^3_1 k^5_2 [4(4s_3-s_2)]\Big\}. \nb\\ 
\eqn

\end{widetext}



\begin{thebibliography}{nbound}

  \bibitem{KDM} L.M. Krauss, S. Dodelson, and S. Meyer, Science {\bf 328}, 989 (2010);
J.  Garcia-Bellido, Prog. Theor. Phys. Suppl. {\bf 190}, 322 (2011). 

 
\bibitem{seljak}
U. Seljak and M. Zaldarriaga, Phys. Rev. Lett. {\bf 78}, 2054 (1997); M. Kamionkowski, A. Kosowsky, and A. Stebbins, {\em ibid.},  {\bf 78}, 2058 (1997);
J. Bock et al., \emph{Task Force on Cosmic Microwave Background Research}, arXiv:astro-ph/0604101.


\bibitem{lue} A. Lue, L. Wang, and M. Kamionkowski, Phys. Rev. Lett. {\bf 83}, 1506 (1999); N. Seto and A. Taruya, {\em ibid.}, {\bf 99}, 121101 (2007).


\bibitem{saito}
S. Saito, K.Ichiki and A.Taruya, JCAP,  {\bf 09}, 002 (2007); C. Bischoff {\em et. al.}, the Quiet Collaboration, Astrophys. J. {\bf 741}, 111 (2011).


\bibitem{kamionkowski}
V. Gluscevic and M. Kamionkowski, Phys. Rev. D{\bf 81}, 123529 (2010).

\bibitem{soda}
T. Takahashi and J. Soda, Phys. Rev. Lett. {\bf 102}, 231301 (2009).

\bibitem{WWZZ} A. Wang, Q. Wu, W. Zhao, and T. Zhu, Phys. Rev. D{\bf 87}, 103512 (2013).


\bibitem{Horava} P. Ho\v{r}ava,  Phys. Rev. D {\bf 79}, 084008 (2009). 



\bibitem{parity} J.M. Maldacena and G.L. Pimentel, JHEP {\bf 09}, 045 (2011). 

\bibitem{parity-soda}J. Soda, H. Kodama and M. Nozawa, JHEP {\bf 08}, 067 (2011). 

\bibitem{parity2} M. Shiraishi, D. Nitta, and S. Yokoyama, Prog. Theor. Phys. 126, 937 (2011).


\bibitem{ZWWS} T.  Zhu, Q. Wu, A. Wang,  and F.-W.  Shu,  Phys. Rev. D {\bf 84}, 101502 (R) (2011).


\bibitem{ZSWW}  T.  Zhu, F.-W.  Shu, Q. Wu, and A. Wang,   Phys. Rev. D {\bf 85}, 044053 (2012).

 \bibitem{WangTensor} A. Wang, Phys. Rev. D{\bf 82}, 124063 (2010).


 \bibitem {HWYZ} Y.-Q. Huang,  A. Wang,  R. Yousefi, and T. Zhu,  Phys. Rev. D{\bf 88}, 023523  (2013).

\bibitem{planck_2013}  P. Ade {\em et. al.}, Planck Collaboration, arXiv:1303.5076.


\bibitem{IKM} K. Izumi, T. Kobayashi, and S. Mukohyama, J. Cosmol. Astropart. Phys. {\bf 10}  (2010) 031.


\bibitem{HW12}  Y.-Q. Huang and  A. Wang, Phys. Rev. D{\bf 86}, 103523 (2012).

\bibitem{Stelle}  K.S. Stelle, Phys. Rev. D{\bf 16},   953 (1977).


 \bibitem{reviews} D. Blas, O. Pujolas, and S.  Sibiryakov,   JHEP {\bf 1104}, 018 (2011); 
                                  P. Ho\v{r}ava, Class. Quantum Grav.  {\bf 28},   114012 (2011); 
		               T. Clifton, P.G. Ferreira, A. Padilla,  and C. Skordis,  Phys. Rep.  {\bf513}, 1 (2012). 
				
				
\bibitem{cosmology}S. Mukohyama, Class. Quantum Grav.  {\bf 27}, 223101 (2010).

 


\bibitem{HMT} P. Ho\v{r}ava and C.M. Melby-Thompson, Phys. Rev. D {\bf 82}, 064027 (2010); 
A. Wang and Y. Wu,   Phys. Rev. D {\bf 83}, 044031 (2011); 
A.M. da Silva,  Class. Quan. Grav.   {\bf 28}, 055011 (2011). 



 
  \bibitem{ADM}   R. Arnowitt, S. Deser, and C.W. Misner, Gen. Relativ. Grav. {\bf 40}, 1997 (2008);
                            C.W. Misner, K.S. Thorne, and J.A. Wheeler, {\em Gravitation}  
                            (W.H. Freeman and Company, San Francisco, 1973), pp.484-528.

\bibitem{Porrati}    M. Porrati,   arXiv:hep-th/0409210.
 


  \bibitem{martin}
J. Martin and R. Brandenberger, Phys. Rev. D{\bf 63}, 123501 (2001).


\bibitem{MB03}
J. Martin and R. Brandenberger, Phys. Rev. D{\bf 68}, 063513 (2003); 
R. Brandenberger and J. Martin, Phys. Rev. D{\bf 71}, 023504 (2005).

                          
\bibitem{in-in} J. Maldacena,  JHEP,  {\bf 05}, 013 (2003); 
				S. Weinberg,   Phys. Rev. D{\bf 72}, 043514 (2005). 

                                 
\bibitem{maldacena}J. Maldacena, JHEP 05 (2003) 013.

\bibitem{Gao-PRL} X. Gao, T. Kobayashi, M. Yamaguchi, and J. Yokoyama, Phys. Rev. Lett. {\bf 107}, 211301 (2011). 




 
\end{thebibliography}
\end{document}